\documentclass[aps,prb,%
twocolumn,
showpacs,floatfix,superscriptaddress]{revtex4}

\usepackage{amsmath,amssymb}
\usepackage[dvips]{graphics}
\usepackage{epsfig}
\usepackage{times}

\begin{document}
\title{Poissonian tunneling through an extended impurity in the quantum Hall effect}
\author{D. Chevallier}
\affiliation{Centre de Physique Th\'eorique, UMR6207, Case 907, Luminy, 13288 Marseille Cedex 9, France}
\affiliation{Universit\'e de la M\'editerran\'ee, Case 907, 13288 Marseille Cedex 9, France}
\author{J. Rech}
\affiliation{Centre de Physique Th\'eorique, UMR6207, Case 907, Luminy, 13288 Marseille Cedex 9, France}
\author{T. Jonckheere}
\affiliation{Centre de Physique Th\'eorique, UMR6207, Case 907, Luminy, 13288 Marseille Cedex 9, France}
\author{C. Wahl}
\affiliation{Universit\'e de la M\'editerran\'ee, Case 907, 13288 Marseille Cedex 9, France}
\author{T. Martin}
\affiliation{Centre de Physique Th\'eorique, UMR6207, Case 907, Luminy, 13288 Marseille Cedex 9, France}
\affiliation{Universit\'e de la M\'editerran\'ee, Case 907, 13288 Marseille Cedex 9, France}

\begin{abstract}
We consider transport in the Poissonian regime between edge states in the quantum Hall effect. The backscattering potential is assumed to be arbitrary, as it allows for multiple tunneling paths. We show that the Schottky relation between the backscattering current and noise can be established in full generality: the Fano factor corresponds to the electron charge (the quasiparticle charge) in the integer (fractional) quantum Hall effect, as in the case of purely local tunneling.  We derive an analytical expression for the backscattering current, which can be written as that of a local tunneling current, albeit with a renormalized tunneling amplitude which depends on the voltage bias. We apply our results to a separable tunneling amplitude which can represent an extended point contact in the integer or in the fractional quantum Hall effect. We show that the differential conductance of an extended quantum point contact is suppressed by the interference between tunneling paths, and it has an anomalous dependence with respect to the bias voltage.
\end{abstract}

\maketitle

\section{Introduction}

Electrons confined to two dimensions and subject to a magnetic field perpendicular 
to this plane exhibit the quantum Hall effect.\cite{klitzing} For sufficiently clean samples and strong fields, 
the excitations of this non trivial state of matter bear fractional charge and statistics: this is the regime
of the fractional quantum Hall effect\cite{laughlin} (FQHE). For samples with boundaries, the edge state 
picture\cite{buttiker_edge,wen} has been quite useful to capture the essential physics.
Over the last two decades, theory\cite{kane_fisher_noise,chamon_freed_wen,saleur} 
and experiment\cite{depicciotto,saminadayar} have addressed the issue of non equilibrium 
transport through quantum Hall bars: when a voltage bias is imposed between two edges a tunneling current transmits quasiparticles 
from one edge state to the other.    
A fundamental property of transport in the tunneling regime is the fact that the Fano factor
-- the ratio between the zero frequency noise and the tunneling current -- should 
correspond to the charge of the carriers which tunnel.\cite{martin_houches,blanter_buttiker}
In the FQHE, for a local, weak impurity potential,
Luttinger Liquid theory predicts\cite{kane_fisher_noise,chamon_freed_wen} that the 
Fano factor corresponds to the effective charge of the quasiparticles.  
Moreover the backscattering current ($I_B$)- voltage ($V$) characteristic has a power law dependence
which is anomalous. Experiments have confirmed the prediction on the Fano factor, 
but the precise theory-experiment correspondence with the $I_B(V)$ characteristics 
remains elusive. 
 
Existing theoretical models have focused mostly on the backscattering associated 
with a single local impurity, which connects a single scattering location on each edge state.
Extensions describing arrays of scattering locations have been considered in Ref.~\onlinecite{sondhi,jolicoeur}. 
However, in practice, quantum point contacts (QPC) consist of electrostatic gates which are 
placed ``high'' above the two dimensional electron gas. In this situation it is quite 
unlikely that the backscattering potential is purely local, and a proper theoretical 
description should take into account multiple tunneling paths. The purpose of the present 
work is to derive the Fano factor for an arbitrary weak backscatterer which takes into 
account such multiple tunneling paths. We show via an analytical argument that the Fano factor remains 
unchanged. Nevertheless, such multiple scattering paths lead to interference phenomena
and the $I_B(V)$ is strongly modified when the impurity has an ``extended'' character.
Analytical expressions are subsequently obtained for $I_B(V)$. We apply our results to a 
quantum point contact which has a characteristic width $\xi$ to illustrate our results.

The paper is organized as follows: in Sec. \ref{model} we introduce the model and the general backscattering Hamiltonian. The current and noise are computed in Sec. \ref{current_and_noise}, and we apply our results to a separable tunneling amplitude in Sec. \ref{application} in order to describe transport through an extended QPC. We conclude in Sec. \ref{conclusion}.   

\section{Model}
\label{model}

\begin{figure}[t]
	\centering
		\includegraphics[width=8.6cm]{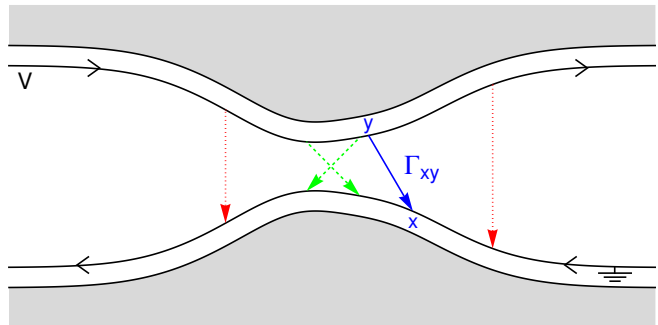}
	\caption{(color online) Description of an arbitrary extended scatterer: 
 the tunneling amplitude from position $y$ on the right propagating edge
  to position $x$ on the left propagating edge is $\Gamma_{xy}e^{i\delta_{xy}}$.
 We show an arbitrary tunneling process (blue) as well as lateral contributions 
 and crossed contributions (respectively red-dotted and green-dashed, see text for details).}
	\label{fig1}
\end{figure}

The typical set up is depicted in Fig.~\ref{fig1}: two counter propagating edge states 
are put in contact by a scattering region. Tunneling of quasiparticles is likely to occur 
in the regions where the two edges are close to each other, but it is plausible that
longer tunneling paths involving different positions on the top and bottom edges
have also to be taken into account. 

We use the Tomonaga-Luttinger formalism to describe the 
right and left moving chiral excitations.
In the absence of tunneling between the two edges, 
the Hamiltonian reads:
\begin{equation}
H_{0}=\frac{v_{F}\hbar}{4\pi}\sum_{r}\int ds(\partial_{s}\phi_r)^2~, 
\end{equation}
with $r=R,L$ for right and left movers. Here $\phi_{r}$ denotes the
bosonic chiral field of each edge.
Correspondingly, we introduce the quasiparticle operator:
\begin{equation} 
\Psi_{r}=\frac{1}{\sqrt{2\pi a}}e^{irk_F x}e^{i\sqrt{\nu}\phi_r(x,t)}
\end{equation}
where $a$ is a short distance cutoff and $\nu$ is the filling factor ($\nu^{-1}$ is an odd integer to describe Laughlin fractions).

Here, we focus  on the weak backscattering regime, but results for strong backscattering
regime can be trivially obtained using the duality transformation \cite{kane}.
The most general Hamiltonian which describes
the backscattering of quasiparticles (with charge $e^*=\nu e$) from the top (right moving) edge to the bottom
(left moving) edge with multiple tunneling paths is described by the Hamiltonian:
\begin{eqnarray}
H_B(t)=\int~dx~ dy \sum_{\varepsilon} [\Gamma_{xy}e^{i\delta_{xy}}\Psi_R^\dag(y,t)\Psi_L(x,t)]^{(\varepsilon)}~,
\end{eqnarray}
where the notation $\epsilon=\pm$ leaves an operator unchanged ($\epsilon=+$) 
or specifies its Hermitian conjugate ($\epsilon=-$). 
Here $\Gamma_{xy}e^{i\delta_{xy}}$ ($\Gamma_{xy}$ is real) is a tunneling amplitude for scattering from point $x$ of the left-moving edge to point $y$ of the right-moving one. An additional phase factor $i\omega_0 t$ ($\omega_0\equiv e^*V/\hbar$) is added to this tunneling amplitude;
it arises from the Peierls substitution in order to take into account the source drain voltage.

A purely local scatterer at $x=0$ corresponds to the choice $\Gamma_{xy}=\Gamma_0\delta(x)\delta(y)$.
In Ref.~\onlinecite{jolicoeur}, the authors considered a point contact over a finite region of space 
with $\Gamma_{xy}= \Gamma_L(x)  \delta (x-y)$. In what follows, 
we label such contributions of $ \Gamma_{xy}$ as ``lateral'' contributions these are indicated in 
red in Fig.~\ref{fig1}. In such lateral contributions, the tunneling Hamiltonian contains rapid 
oscillations due to the presence of the phase factor $i2k_Fx$.
Another contribution is the case of so called ``crossed'' 
contributions $\Gamma_{xy} = \Gamma_C(x) \delta (x+y)$  (green line in Fig.~\ref{fig1}) for which such 
$2k_F$ oscillations are absent. Nevertheless, the crossed contribution can be shown to also 
exhibit oscillations, but on a much longer 
lengthscale $2\pi v_F/\omega_0$ (see below).

\section{Poissonian Current and Noise}
\label{current_and_noise} 

The local backscattering current is deduced from the backscattering Hamiltonian:
\begin{eqnarray}
I_{Bxy}(t)=\frac{ie^*}{\hbar}\sum_{\varepsilon}\varepsilon \Gamma_{xy}e^{i\epsilon\delta_{xy}}e^{i\epsilon\omega_0 t}[\Psi_R^\dag(y,t)\Psi_L(x,t)]^{(\varepsilon)}
~.\end{eqnarray}
The partial average current is computed using the Keldysh formalism:
\begin{equation}
\langle I_{Bxy}(t)\rangle=\frac{1}{2} \sum_{\eta=\pm} \langle T_K I_{Bxy}(t^\eta)e^{\frac{1}{i\hbar}\int_Kdt' H_B(t')}\rangle
\end{equation}
where $\eta$ identifies which part of the Keldysh contour is chosen.
For the Poissonian limit of weak backscattering, the exponential is expanded to first order in $H_B$.
The definition of the partial, symmetrized real time noise correlator in the Heisenberg  representation reads:
\begin{eqnarray}
S_{xyx'y'}(t,t')&=& \langle I_{Bxy}(t)I_{Bx'y'}(t')\rangle/2 
\nonumber\\&~&~
+ \langle I_{Bx'y'}(t')I_{Bxy}(t)\rangle/2 
\nonumber\\&~&~-\langle I_{Bxy}(t)\rangle \langle I_{Bx'y'}(t') \rangle 
\nonumber\\
&=&\sum_{\eta=\pm} \langle I_B(t^\eta)I_B(t'^{-\eta})\rangle/2 ~,
\end{eqnarray}
where the second equality is written in the interaction representation, to the same (2nd) order in $\Gamma_{xy}$ as for the current
(the product of current averages contributes to higher order in $H_B$).

Inserting the expression of the quasiparticle operators in the current and using so called ``quasiparticle conservation'' we obtain:
\begin{eqnarray}
\langle I_{Bxy}(t)\rangle
&=&\frac{e^*}
{8\pi^2a^2 \hbar^2}\Gamma_{xy}\int dx'dy'\Gamma_{x'y'}
\sum_{\eta\eta'\epsilon} \epsilon \eta'
\int dt'
\nonumber\\&~&~
e^{i\epsilon(\omega_0(t-t')-k_F (x+y-x'-y')+\delta_{xy}-\delta_{x'y'})}
\nonumber\\
&~&~
\langle T_K e^{-i\epsilon\sqrt{\nu}\phi_R(y,t^\eta)} e^{i\epsilon\sqrt{\nu}\phi_R(y',t'^{\eta'})} \rangle
\nonumber\\&~&~
\langle T_K e^{i\epsilon\sqrt{\nu}\phi_L(x,t^\eta)} e^{-i\epsilon\sqrt{\nu}\phi_L(x',t'^{\eta'})} \rangle ~,
\end{eqnarray}
where all integrals in the remainder of this paper (unless specified) run from $-\infty$ to $+\infty$.
Bosonised expressions of the field operators are inserted in the time ordered products, which in turn 
are expressed in terms of the chiral Green's functions:
\begin{equation}
\langle T_K e^{\alpha\phi_{L,R}(x,t^\eta)}e^{\beta\phi_{L,R}(0,0^{\eta'})}\rangle
=e^{\alpha\beta G_{L,R}^{\eta\eta'}(t\mp x/v_F)}
\end{equation}
for $\alpha=-\beta$. At zero temperature:
\begin{align} 
&G_{L,R}^{\eta-\eta}(t)=-\ln (1-i\eta v_F t/a)\\
&G_{L,R}^{\eta\eta}(t)=-\ln (1+i\eta v_F |t|/a).
\end{align}
Performing a change of variable on times, this gives
\begin{eqnarray}
\langle I_{Bxy}(t)\rangle
&=&-\frac{ie^*}{4\pi^2a^2 \hbar^2}\Gamma_{xy}\int dx' dy' \Gamma_{x'y'}
\sum_{\eta\eta'}\eta'\int d\tau
\nonumber\\
&~&~
\times \sin(\omega_0\tau-k_{F+}(x-x')-k_{F-}(y-y')
\nonumber\\
&~&~~~~~
+\delta_{xy}-\delta_{x'y'})
\nonumber\\
&~&~
\times e^{\nu G^{\eta\eta'}(\tau+\frac{z}{v_F})}e^{\nu G^{\eta\eta'}(\tau-\frac{z}{v_F})}~,
\label{IBxy_intermed}
\end{eqnarray}
where $z\equiv (x-x'+y-y')/2$ and $k_{F\pm}\equiv k_F\pm\omega_0/2v_F$. Similarly, for the real time noise correlator, we find: 
\begin{eqnarray}
S_{xyx'y'}(t,t')
&=&\frac{e^*}{4\pi^2a^2 \hbar^2}\Gamma_{xy}\Gamma_{x'y'}
\sum_{\eta}
\nonumber\\
&~&
\cos(\omega_0(t-t')-k_{F+}(x-x')-k_{F-}(y-y')
\nonumber\\
&~&~~~~~
+\delta_{xy}-\delta_{x'y'})
\nonumber\\
&~&~
e^{\nu G^{\eta,-\eta}(t-t'+\frac{z}{v_F})}e^{\nu G^{\eta,-\eta}(t-t'-\frac{z}{v_F})}~.
\end{eqnarray}

We now focus on the total backscattering current and noise which sum all possible paths:
\begin{eqnarray}
\langle I_{BT}\rangle&\equiv& \int dx~dy~\langle I_{Bxy}(t)\rangle\\
S_T(t,t')&\equiv& \int dx~dy\int dx'~dy'~S_{xyx'y'}(t,t').
\end{eqnarray}
Noticing that the sine function in (\ref{IBxy_intermed}) is odd under the transformation $\tau\to-\tau$, $x\leftrightarrow x'$, $y\leftrightarrow y'$, only the contribution $\eta=-\eta'$ remains. Moreover, the contributions $\eta=\pm$ give the same result, therefore:
\begin{align}
\langle I_{BT}\rangle =& -\frac{ie^*}{2\pi^2a^2 \hbar^2}\int dx dy \int dx' dy' ~\Gamma_{xy}\Gamma_{x'y'}
\nonumber\\
&~
\cos(k_{F+}(x-x')+k_{F-}(y-y')-\delta_{xy}+\delta_{x'y'})
\nonumber\\
&~\int d\tau \frac{\sin(\omega_0 \tau)}{\left[1-i\frac{v_F}{a}(\tau+\frac{z}{v_F})\right]^\nu \left[1-i\frac{v_F}{a}(\tau-\frac{z}{v_F})\right]^\nu}~.
\label{current_total}
\end{align}
Next, because we are interested in the total noise at zero frequency, we perform the integral $\tilde{S}_T$ 
of $S_T(t,t')$ over the variable $t-t'$.
Exploiting once again the parity properties of the Green's functions with the summation over $\eta$, one obtains:
\begin{align}
\tilde{S}_T
=&\frac{e^{*2}}{2\pi^2a^2\hbar^2}\int dx dy  \int dx' dy' ~\Gamma_{xy}\Gamma_{x'y'}
\nonumber\\
&~
\cos(k_{F+}(x-x')+k_{F-}(y-y')-\delta_{xy}+\delta_{x'y'})
\nonumber\\
&~\int d\tau \frac{\cos(\omega_0 \tau)}{\left[1-i\frac{v_F}{a}(\tau+\frac{z}{v_F})\right]^\nu \left[1-i\frac{v_F}{a}(\tau-\frac{z}{v_F})\right]^\nu}~.
\label{noise_total}
\end{align}
We notice that both contributions for the current and noise involve the integrals:
\begin{equation}
J_{\pm}(\omega_0,z)=\int d\tau \frac{e^{\pm i |\omega_0| \tau}}{\left[1-i\frac{v_F}{a}(\tau+\frac{z}{v_F})\right]^\nu \left[1-i\frac{v_F}{a}(\tau-\frac{z}{v_F})\right]^\nu}
\end{equation}
(indeed, $\langle I_{BT} \rangle$ involves the combination $(J_+-J_-)/2i$ while $S_T$ contains $(J_++J_-)/2$).
These integrals can be expressed in terms of the variable $w=\omega_0 \tau \equiv w'+i w''$.
The integrand has a branch cut at the location ($w'=\pm z |\omega_0|/v_F$, $w''\leq -a|\omega_0|/v_F$). We can extend the integral $J_{\pm}(\omega_0,z)$ as a closed contour in the upper half plane.
We notice that there are no poles or branch cuts for $J_{+}(\omega_0,z)$ in this plane, 
so Cauchy theorem tells us that the integral from $-\infty$ to $+\infty$ is zero.
Only $J_{-}(z)$ contributes.

Substituting this result back into Eqs.~(\ref{current_total}) and (\ref{noise_total}), one readily sees that the ratio of the zero-frequency noise to the tunneling current simplifies 
\begin{equation}
\frac{\tilde{S}_T}{\langle I_{BT}\rangle} = e^{*} ~,
\end{equation}
independently of the details of the tunneling amplitude. Thus, although the extended character of the contact can dramatically affect the current and noise, the Fano factor is left unchanged, taking the expected value for a Poissonian process.

We now turn to the analytical derivation of the backscattering current.
First we notice that at $z=0$, analytical expressions are available: 
\begin{equation}
J_-(\omega_0,0)\equiv \frac{2\pi}{{\bf \Gamma}(2\nu)}\left(\frac{a}{v_F}\right)^{2\nu} |\omega_0|^{2\nu-1}
\end {equation}
where ${\bf \Gamma}$ is the gamma function. For finite $z$, one can express $J_-(\omega_0, z)$ in terms of its $z=0$ counterpart\cite{sondhi} as
\begin{eqnarray}
J_-(\omega_0,z)&=&\frac{2\pi}{{\bf \Gamma}^2(\nu)} \left(\frac{a}{v_F}\right)^{2\nu}
\int_0^{|\omega_0|} d\omega' \omega'^{\nu-1} \nonumber \\
& &\times (|\omega_0|-\omega')^{\nu-1} e^{i(2\omega'-|\omega_0|)\frac{z}{v_F}}
\nonumber\\
&=&~J_-(\omega_0,0)H_\nu \left(|\omega_0| \frac{z}{v_F}\right)
\end{eqnarray}
with
\begin{equation}
H_\nu(y)\equiv \sqrt{\pi}\frac{{\bf \Gamma}(2\nu)}{{\bf \Gamma}(\nu)}\frac{J_{\nu-1/2}(y)}{(2y)^{\nu-1/2}}
\end{equation}
where $J_{\nu-1/2}(y)$ is the Bessel function of the first kind.

This allows to rewrite the total backscattering current in the same form as for a purely 
local point contact:
\begin{eqnarray}
\langle I_{BT}\rangle =\frac{e^*|\Gamma_{\rm eff}(\omega_0)|^2}{2\pi a^2 \hbar^2{\bf
\Gamma}(2\nu)}
\left(\frac{a}{v_F}\right)^{2\nu}|\omega_0|^{2\nu-1} {\rm Sgn}(\omega_0)
~,
\end{eqnarray}
where:
\begin{align}
|\Gamma_{\rm eff}(\omega_0)|^2 &=\int dx dy \int dx' dy' ~\Gamma_{xy} \Gamma_{x'y'} H_\nu\left(\frac{|\omega_0| z}{v_F}\right)
\nonumber\\
&~\cos(k_{F+}(x-x')+k_{F-}(y-y')-\delta_{xy}+\delta_{x'y'}) ~.
\label{gamma_eff}
\end{align}
Note that in the purely local case $\left|\Gamma_{\rm eff}\right|^2=\left|\Gamma_0\right|^2$. In the general case the effective tunneling amplitude has a non trivial dependence on the potential bias which triggers a deviation from the power law dependence $I_B\sim \omega_0^{2\nu-1}$.

\section{Application to a separable tunneling amplitude}
\label{application}   

Assuming a separable form for the local 
tunneling amplitude, one can write:
\begin{equation}
\Gamma_{xy} e^{i\delta_{xy}}=\Gamma_0 g_+(x+y) g_-(x-y)
\end{equation}
where  $g_{\pm}$ are functions which are typically maximal around zero and 
which decrease otherwise. The role of $g_+$ is to specify the 
average location of the impurity, while $g_-$ expresses the fact 
that long tunneling paths carry less weight than short ones. 
Under this assumption, it becomes possible to fully decouple in Eq.~(\ref{gamma_eff}) the integrals over $x-y$ and $x'-y'$ from the ones over $x+y$ and $x'+y'$. As a result, one can recast the effective tunneling probability $|\Gamma_{\rm eff}(\omega_0)|^2$ as the product of a crossed and a lateral contribution, namely
\begin{equation}
\left| \frac{\Gamma_{\rm eff}(\omega_0)}{\Gamma_0} \right|^2 = \left| \frac{\Gamma_{\rm eff}^C(\omega_0)}{\Gamma_0} \right|^2 \times \left| \frac{\Gamma_{\rm eff}^L(\omega_0)}{\Gamma_0} \right|^2
\end{equation}
where these two terms are defined as
\begin{align}
|\Gamma_{\rm eff}^C(\omega_0)|^2 &=\int dx dx' ~|\Gamma_C (x)|~| \Gamma_C (x') | 
\nonumber\\
&~\times \cos \left( \frac{\omega_0}{v_F}(x-x')-\delta_C (x)+\delta_C (x') \right) \\
|\Gamma_{\rm eff}^L(\omega_0)|^2 &=\int dx dx' ~|\Gamma_L (x)|~| \Gamma_L (x')| H_\nu\left(|\omega_0| \frac{x-x'}{v_F}\right)
\nonumber\\
&~\times \cos \left( 2 k_F (x-x')-\delta_L (x)+\delta_L (x') \right)
\end{align}
with $\Gamma_C (x) = |\Gamma_C (x)| e^{i\delta_C (x)} = 2 \Gamma_0 g_-(2x)$ and $\Gamma_L (x) = |\Gamma_L (x)| e^{i\delta_L (x)} = \Gamma_0 g_+(2x)$. Surprisingly, the crossed contribution does not depend explicitly on the filling factor $\nu$, but only implicitly through $\omega_0$. 

\begin{figure}[t]
	\centering
		\includegraphics[width=6.5cm]{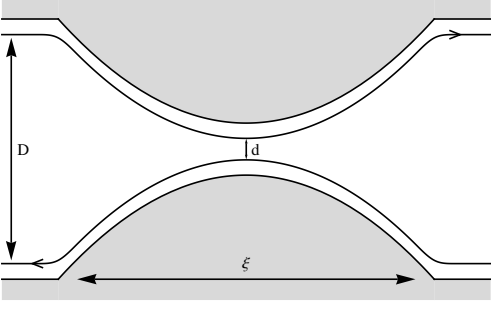}
	\caption{A ``Parabolic Quantum point contact": the edge states follow a parabolic profile of width $\xi$
	with a minimum distance $d$ between them at the center of the QPC. }
	\label{fig:parabolic_QPC}
\end{figure}

We now use our previous calculations to study the effective tunneling amplitude and the differential conductance as a function of two parameters: the applied voltage $\omega_0$ and the width of the contact region $\xi$. The tunneling amplitude arises from the overlap between states from the top and bottom edges, and as such is proportional to $e^{-(l_{xy}^2/4l_B^2)}$, where $l_B=\sqrt{\hbar/e B}$ is the magnetic length and $l_{xy}$ is the distance between the position $y$ on the top edge and the position $x$ on the bottom edge. Relying on this argument, we choose $\Gamma_{xy}$ to be Gaussian:
\begin{equation}
\Gamma_{xy}=\frac{1}{2\pi\xi_{c}\xi_{l}}e^{-\frac{(x-y)^2}{4\xi_{l}^2}}e^{-\frac{(x+y)^2}{4\xi_{c}^2}}.
\label{gamma_gaussian}
\end{equation}
We believe that this simple choice can represent accurately a large number of geometries. Let us consider for example, a parabolic QPC, where the edge states follow a parabolic profile of width $\xi$ (see Fig.\ref{fig:parabolic_QPC}). To fully describe this geometry, we introduce the minimal distance $d$ between edge states inside the constriction, as well as the width $D$ of the Hall bar, with $D \gg d$. One can show that for such a geometry, a $\Gamma_{xy}$ of the form given in Eq.(\ref{gamma_gaussian}) can be recovered, provided that $\xi^2/(d D) \gg 1$. In this configuration, $\xi_{c}$ is typically constant and equal to the magnetic length $l_B$, and $\xi_l$ is set by the width of the contact region, $\xi_{l}=\xi l_B/\sqrt{dD}$. While these relations between the characteristic scales $\xi_{l,c}$ and the geometrical parameters defining the QPC are specific to the parabolic case, the Gaussian profile introduced in Eq.(\ref{gamma_gaussian}) is general enough to account for many other situations and the results presented below are believed to be valid beyond this simple parabolic picture.

Inserting the expression (\ref{gamma_gaussian}) for the tunneling amplitude in Eq.~(\ref{gamma_eff}) and performing the integrals over space variables we obtain:  
\begin{align}\label{Gaussianform}
\left|\Gamma_{\rm eff}\right|^2&=\frac{\Gamma_0^2}{2\sqrt{2\pi}\xi_{l}}e^{-\frac{\xi_{c}^2\omega_0^2}{2v_F^2}}\\\notag
&\times\int dx \textrm{cos}(k_F x)e^{-\frac{x^2}{8\xi_{l}^2}}H_{\nu}\left(\frac{\left|\omega_0\right|x}{2 v_F}\right).
\end{align}
An essential dimensionless parameter in the above integral is $k_F \xi_{l}$. The Fermi momentum $k_F$ inside the constriction can be estimated as $k_F \simeq d/2l_B^2$.\cite{kim_fradkin2003} As $d$ should be of the order of $l_B$, this gives $k_F \sim 1/l_B$. For a value of $B$ typical of 2d electron gas QHE ($B \sim 6$T), one has $l_B \simeq 10$ nm, and thus $k_F \xi_{l}$ range from 1 to 10 for values of $\xi_{l}$ between $\sim 10$ and $\sim 100$ nm.
 
\begin{figure}[h]
	\centering
		\includegraphics[width=8cm]{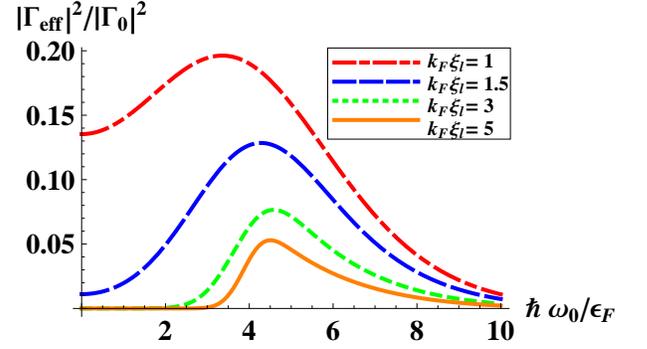}
	\caption{(color online) $\left|\Gamma_{\rm eff}\right|^2$ normalized by $\Gamma_0^2$ as a function of the applied voltage $\hbar\omega_0/\epsilon_F$, for $k_F \xi_c=0.5$ and various values of $k_F\xi_{l}$ in the integer quantum Hall effect ($\nu=1$).}
	\label{fig:Courbe1}
\end{figure}

\begin{figure}[h]
	\centering
		\includegraphics[width=8cm]{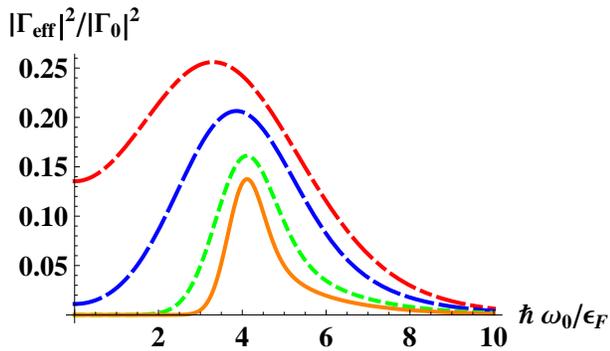}
	\caption{(color online) $\left|\Gamma_{\rm eff}\right|^2$ normalized by $\left|\Gamma_0\right|^2$ as a function of the applied voltage $\hbar\omega_0/\epsilon_F$ for $k_F \xi_c=0.5$ and various values of $k_F\xi_{l}$ in the fractional quantum Hall effect with a filling factor $\nu=1/3$ (The values chosen for $k_F\xi_{l}$ are the same as in Fig.~\ref{fig:Courbe1}).}
	\label{fig:Courbe2}
\end{figure}

In Fig.~\ref{fig:Courbe1} we see that in the integer quantum Hall regime ($\nu=1$) the effective amplitude of tunneling decreases when we increase $k_F\xi_{l}$. Moreover, $\left|\Gamma_{\rm eff}\right|^2$ has a non-monotonic behavior in voltage with a maximum around $\hbar\omega_0/\epsilon_F=4$, which is in sharp contrast with the constant value expected in the purely local case. Fig.~\ref{fig:Courbe2} shows the behavior of $\left|\Gamma_{\rm eff}\right|^2$ in the regime of the fractional quantum Hall effect ($\nu=1/3$). This behavior is similar to that of the integer regime, only the curves are more peaked around the maximum.
 
\begin{figure}[h]
	\centering
		\includegraphics[width=8cm]{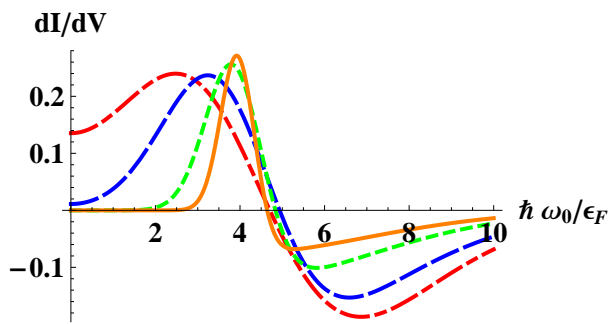}
	\caption{(color online) Differential conductance normalized by $\left.dI/dV\right|_{(0)}$ as a function of $\hbar\omega_0/\epsilon_F$ for $k_F \xi_c=0.5$ and various values of $k_F\xi_{l}$ in the integer quantum Hall effect ($\nu=1$) (The values chosen for $k_F\xi_{l}$ are the same as in Fig.~\ref{fig:Courbe1}).}
	\label{fig:DIDV1}
\end{figure}

We now turn to the computation of the differential conductance $dI/dV$ which, for convenience, is normalized in all plots to the following value:
\begin{equation}
\left.dI/dV\right|_{(0)}=\frac{(e^*)^2}{2\pi a^2 \hbar^{2\nu+1} {\bf
\Gamma}(2\nu)}
\left(\frac{a}{v_F}\right)^{2\nu}\epsilon_F^{2\nu-2}\Gamma_0^2.
\end{equation}
In Fig.~\ref{fig:DIDV1}, we plot $dI/dV$ in the integer quantum Hall regime ($\nu=1$) as a function of the applied voltage for various values of $k_F\xi_{l}$. 
While in the purely local case, the differential conductance is expected to be constant, here it shows a peaked structure, and becomes negative shortly after reaching its maximum value. 
As $k_F\xi_{l}$ increases, a threshold appears at low voltage, below which the differential conductance takes vanishingly small values. This suppression is reminiscent of what has been observed in previous theoretical\cite{jolicoeur} and experimental\cite{kang} works, in the context of quantum Hall line junctions, where current suppression below a given threshold for long barriers was related to momentum conservation.

\begin{figure}[h]
	\centering
		\includegraphics[width=8cm]{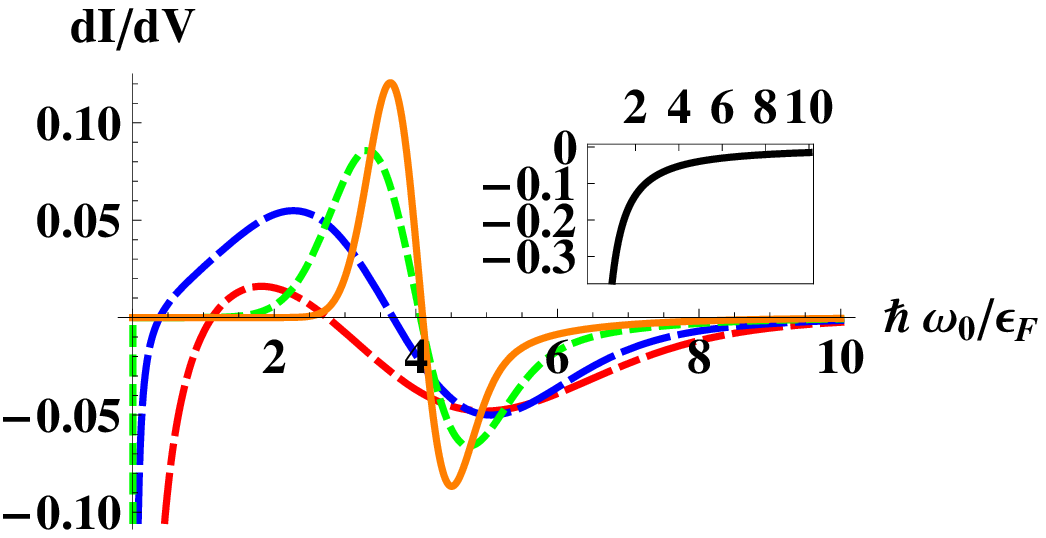}
	\caption{(color online) Differential conductance normalized by $\left.dI/dV\right|_{(0)}$ as a function of $\hbar\omega_0/\epsilon_F$ for $k_F \xi_c=0.5$ and various values of $k_F\xi_{l}$ in the fractional quantum Hall effect with a filling factor $\nu=1/3$ (The values chosen for $k_F\xi_{l}$ are the same as in Fig.~\ref{fig:Courbe1}). Inset: Differential conductance in the purely local case.}
	\label{fig:DIDV13}
\end{figure}

In Fig.~\ref{fig:DIDV13}, we plot the differential conductance in the fractional quantum Hall regime ($\nu=1/3$) as a function of the applied voltage for various values of $k_F\xi_{l}$. In the inset we show the purely local case with the characteristic Luttinger divergence at $\omega_0\rightarrow0$. This power-law behavior survives for the extended contact at very low voltage, over a region that shrinks rapidly as $k_F\xi_{l}$ increases. Beyond this low-voltage regime, the differential conductance shows the same kind of peaked structure as in the $\nu=1$ case. Interestingly, for large values of $k_F\xi_{l}$, the behavior of $dI/dV$ obtained for the integer and fractional Hall regimes are very similar, while substantially deviating from their purely local counterparts. 

\section{conclusion}
\label{conclusion}

To summarize, we have studied current and noise in the integer and fractional QHE in the presence of 
a weak arbitrary backscatterer which allows multiple tunneling paths. 
This calculation in the Poissonian limit shows that the Fano factor
corresponds to the charge of the quasiparticles which tunnel from one 
edge to the other. While this could be considered as an expected result, 
no such general derivation was available so far. We have provided an analytical 
derivation of the tunneling current, where we see explicitly that it does not obey the standard power law behavior of the purely local case. 
Results showing the dependence of the effective tunneling amplitude (which enters the backscattering current) and differential conductance
on the extent of the impurity have been illustrated for a symmetric extended point contact.   
A generalization of this work to finite temperatures could be envisioned. Finally, 
the opposite regime of strong backscattering (where the quantum Hall fluid is split in 
two and only electrons can tunnel between the two edges) can be trivially obtained with the 
duality substitution $\omega_0\to eV/\hbar$ and $\nu\to 1/\nu$. 

\acknowledgments
The authors thank D. C. Glattli and P. St\v{r}eda for helpful discussions. This work is supported by the ANR grant ``1 shot'' (2010).

\end{document}